\documentclass[12pt]{JHEP}

\usepackage{amsmath,epsfig}
\usepackage{amssymb,amsfonts}
\usepackage{latexsym}
\relax

\def\hri#1#2{\href{http://arxiv.org/abs/#1}{[ArXiv:#1]#2}}
\def\hre#1#2{\href{http://arxiv.org/abs/#1/#2}{[ArXiv:#1/#2]}}

\def\be{\begin{equation}}
\def\ee{\end{equation}}
\def\bea{\begin{eqnarray}}
\def\eea{\end{eqnarray}}

\newcommand\fverb{\setbox\pippobox=\hbox\bgroup\verb}
\newcommand\fverbdo{\egroup\medskip\noindent%
                        \fbox{\unhbox\pippobox}\ }
\newcommand\fverbit{\egroup\item[\fbox{\unhbox\pippobox}]}

\newcommand{\la}{\lambda}

\newcommand{\bear}{\begin{eqnarray}}

\newcommand{\eear}{\end{eqnarray}}

\newbox\pippobox

\def\ie{{\it i.e.~}}
\def\lab{\label}
\def\6{\partial}
\def\f{\Phi}
\def\a{\alpha}

\def\half{\frac12}

\def\le{\left}
\def\ri{\right}
\def\cO{{\cal O}}

\def\C0{{\bf C_0}}
\def\Y0{{\bf Y_0}}
\def\G0{{\bf G_0}}
\def\e{\epsilon}
\def\m{\mu}

\def\sq
\def\a{\alpha}
\def\b{\beta}
\def\l{\lambda}

\def\tr{{\rm Tr}}

\def\eps{\epsilon}
\def\cG{{\cal G}}
\def\La{\Lambda}

\def\be{\begin{equation}}
\def\ee{\end{equation}}
\def\bea{\begin{eqnarray}}
\def\eea{\end{eqnarray}}

\def\ie{{\it i.e.~}}
\def\lab{\label} 
\def\6{\partial}
\def\f{\Phi}
\def\a{\alpha}

\def\half{\frac12}

\def\le{\left}
\def\ri{\right}
\def\cO{{\cal O}}

\def\e{\epsilon}
\def\m{\mu}

\def\sq
\def\a{\alpha}
\def\b{\beta}
\def\l{\lambda}
\def\La{\Lambda}

\def\tr{{\rm Tr}}

\def\eps{\epsilon}

\def\F{\Phi}
\def\la{\langle}
\def\ra{\rangle}
\def\g{\gamma}
\def\go{\g_{00}}
\def\gi{\g_{ii}}

\def\cG{\mathcal{G}}

\title{Deconfinement and Thermodynamics in 5D Holographic Models of QCD}
\author{U. G{\"u}rsoy,\\
Institute for Theoretical Physics, Utrecht University;
Leuvenlaan 4, 3584 CE Utrecht, The Netherlands.}

\abstract{We review 5D holographic approaches to finite temperature QCD.
Thermodynamic properties of the ``hard-wall" and the ``soft-wall"
models are derived. Various non-realistic features in these models
are cured by the set-up of improved holographic QCD, that we
review here. }


\begin{document}

\def\g{\gamma}
\def\go{\g_{00}}
\def\gi{\g_{ii}}

\maketitle 

\section{Introduction}

Recent experimental results indicate that the quark-gluon plasma
produced in the heavy-ion collisions stays strongly coupled at
temperatures above deconfinement \cite{RHIC}. Therefore
understanding the nature of QCD matter at high temperature and
density requires non-perturbative techniques. Lattice QCD, being
an intrinsically Euclidean formulation is not well-suited for
calculating certain important dynamical observables such as
transport coefficients or any sort of real-time correlation
functions.

For this reason holographic techniques based on the AdS/CFT
correspondence \cite{AdSCFT}, have recently attracted much
attention in study of such dynamical phenomena. For example the
shear viscosity \cite{shear}, jet quenching parameter \cite{JQP}
and the drag force \cite{drag} has been calculated with better
success than corresponding perturbative findings.

One approach for constructing gravitational backgrounds dual
to QCD-like theories, is to search for deformations of ten
dimensional $AdS_5\times S^5$ background that breaks supersymmetry
and conformality. Such models \cite{SakaiSugimoto} enjoyed success
in reproducing certain IR phenomena but they also bear some
non-realistic features such as presence of KK modes arising
from the extra dimensions. Another, more phenomenological approach
\cite{PS}, instead of attempting at deriving QCD from
fundamentals of 10D critical string theory, aims at deriving a 5D gravitational background
 from the basic requirements of QCD.
 This idea goes under the name of AdS/QCD\cite{AdSQCD} and also achieved partial success, especially
 in the meson sector.

 Generally, finite temperature in the
 holographic  approach is introduced by compactifying the Euclidean
 time direction with period $1/T$. One such obvious solution is the thermal graviton gas.
Other more non-trivial solutions involve black-holes.
 The black-hole solutions correspond to the deconfined phase of the corresponding gauge theory
 \cite{Witten1}, hence encode physics above the deconfinement transition.
 The purpose of this paper is to review the thermodynamics in the aforementioned 5D models.

   In the next  section, we review the thermodynamic properties of the AdS/QCD models
based on the hard-wall (HW) and the soft-wall (SW) geometries. By
extending the analysis of \cite{Herzog}, we compute quantities
such as the energy, entropy and speed of sound as functions of T
and compare them with the expectations from the lattice. Unlike
the HW, the SW model shows good agreement with the lattice data.
However, being a non-dynamical model the black-holes in this
geometry do not obey the laws of BH thermodynamics. From a
practical point of view this fact renders computation of certain
quantities like the bulk viscosity ill-defined. Moreover, it does
not give insight in the nature of the deconfinement transition.

In section 3, we study a dynamical model based on dilaton-gravity
\cite{GK,GKN} which is close in many respects to real QCD. This model is based on general expectation of stringy holographic QCD whose thermodynamic properties were derived in
\cite{GKMN1}. We show that this background solves most of the
problems in the AdS/QCD models at once, sheds light on the role of
the gluon condensate in the phase transition and yields very good
agreement with the lattice data.  In the final section, we
summarize the results, and discuss further directions.

\section{Themodynamics of the AdS/QCD Models}\label{sec2}

\subsection{Hard-wall model}\lab{HW}

The simplest 5D holographic model for QCD is introduced in
\cite{AdSQCD}. The idea is based on the fact that QCD behaves
nearly scale invariant for a wide range of energies ranging from
far UV down to medium energy scales. Thus the authors of
\cite{AdSQCD} proposed a geometrical set-up based on the 5D AdS
space with a cut-off in the deep-interior of the holographic
coordinate $r$. The cut-off is introduced in order to break the
conformal invariance in the IR, and eventually to model color
confinement. The location of the cut-off at $r=r_0$, is dual to
the dynamically generated energy scale of QCD as $\Lambda_{QCD}
\sim 1/r_0$. We shall refer to this solution, as the ``AdS
cavity'' for short.

The model captures many basic features of QCD: one finds a
discrete glueball spectrum by studying the fluctuations of the
metric, an area law for the Wilson loop by studying classical
string embeddings \cite{Wilson}, etc. However, the real success of
the model is in the meson sector and indeed the intention of the authors of \cite{AdSQCD} 
was to apply it there. In all of the 5D models that are
discussed in this paper, the meson sector is generated by
space-filling $D4$ and $\overline{D}4$ branes. The fluctuations of the
brane fields produce the meson spectra. In addition to reproducing
certain generic features such as chiral symmetry breaking,
existence of Nambu-Goldstone fields, Gell-Mann-Oaks-Renner
relation, one finds \%9 agreement with experimental data in
the $1^{\pm}$, $0^{\pm}$ and $1^{++}$ spectra.

Yet, the model is crude in many ways, especially when it comes to the glue sector.  
Running of the gauge coupling
is not taken  into account; not only the electric but also the
magnetic quarks are confined (the 't Hooft loop also exhibits an
area law); there is an ambiguity in computation of the glueball masses and
related to this there is a degeneracy in the $2^{++}$ and $0^{++}$ glueballs
\cite{GKN}; both the glueball and meson spectra are quadratic for
large orbital quantum numbers. This list can be largely expanded,
but here we shall focus our attention on the thermodynamics of the
model and show that the model does not correctly fulfill expectations
for the finite temperature physics either.

\subsection{Thermodynamics of the Hard-Wall}\lab{thermoHW}

Some thermodynamical aspects of the hard-wall model is
investigated in \cite{Herzog}. At finite temperature there are two
competing solutions with the same asymptotics on the boundary: i)
a thermal graviton gas which is just the AdS cavity with compact
Euclidean time of circumference $1/T$ ii) the AdS black-hole
solution with horizon at $r=r_h$. The temperature is related to
the location of horizon as $T = 1/\pi r_h$. As one heats up the
black-hole, the horizon expands and at a particular temperature it
coincides with the IR cut-off $r_h=r_0$.  This is the minimum
temperature for presence of the black-hole inside the AdS cavity
$T_{min} = 1/\pi r_0$. To find out the true minimum of the free
energy, one computes the action evaluated on i) and ii) and then
one takes the difference, see \cite{Herzog} for details.

 Let us define the IR scale as $\Lambda=1/r_0$. Then, the free energy
density\footnote{A word on notation: We shall define the thermodynamic {\em
densities} as the thermodynamic function divided by the volume of
the 3D space $V_3$ times the number of degrees of freedom $N_c^2$.
For example the entropy density is $s=S/(V_3 N_c^2)$.} for the hard-wall model, for $T>T_{min}$
reads\footnote{Our notation for the action reads $S = -M_p^3
N_c^2\int \sqrt{g} (R + V)$. Note that this involves an extra
factor of 2 with respect to \cite{Herzog} where his $\kappa$ is
related to our $M_p$ as $\kappa^{-1} = M_p^{\frac32} N_c$. },
\be\lab{freeHW} f_{HW} = (M_p\ell)^3 \Lambda^4 \le[ 2 - \pi^4
\le(\frac{T}{\Lambda}\ri)^4 \ri]. \ee Here $M_p$ is the Planck
scale and $\ell$ is the AdS radius.  One finds a
confinement-deconfinement phase transition at $T_c = 2^{\frac14} \Lambda/\pi$.
The pressure density is given by $p_{HW} = -f_{HW}$.

In order to compare the analytic results of the hard-wall model
with the lattice data for QCD, one should fix the various energy
scales in the model, \ie $M_p\ell$ and $\La$. The latter is
usually fixed by comparing the vector meson spectrum of the model
with the lattice data\cite{AdSQCD}. One obtains $\La=323$ MeV.
This, in particular yields a transition temperature at $T_c =
122.3$ MeV \cite{Herzog}, (see eq. (\ref{freeHW})).

Fixing the Planck mass is more tricky. The most rigorous way
 is by comparing the high T asymptotics of QCD and the
holographic model. At very high temperatures, the (quenched) QCD
becomes a free gas of gluons with a limit value for the pressure
density $p_{QCD}/T^4\to \pi^2/45$ as $T\to\infty$. In the
hard-wall model, we find from (\ref{freeHW}) that the same
quantity limits to $(M_p\ell)^3\pi^4$. Equating the two yields,
\be\lab{planck} M_p\ell = (45\pi^2)^{-\frac13}.\ee We stress that
{\em this is a universal, model independent way of fixing the
Planck mass:} One obtains the same value for all of the models
discussed in this paper. This is guaranteed to happen quite
generally, if the geometry asymptotes to an AdS black-hole near
the boundary.

\begin{figure}[h]
\centerline{\psfig{file=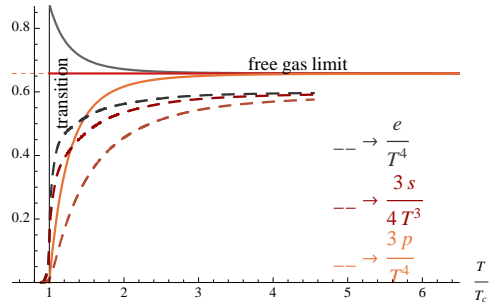,width=2.5in}}
\vspace*{8pt} \caption{Comparison of the energy, entropy and
pressure densities in the HW model with the lattice data of Boyd
et al. (dashed curves).\protect}
\label{fig1}
\end{figure}


Having fixed all of the parameters in the model, we can compare
the thermodynamic functions derived from the HW model with the
lattice results. From (\ref{freeHW}) follow all thermodynamic quantities by 
standard rules.  The entropy density can
either be found by $s = -df/dT$ or by the Bekenstein-Hawking
formula which relates it to the area of the horizon. Eventually,
the energy density follows from $e = f + sT$. All in all, one has,
\be\lab{esHW} s_{HW}  =  4(M_p\ell)^3 \pi^4 T^3,\qquad e_{HW} =
(M_p\ell)^3 \Lambda^4 \le[2+ 3 \pi^4
\le(\frac{T}{\Lambda}\ri)^4\ri], \ee
where $M_p\ell$ is given by (\ref{planck}).
Energy, entropy and pressure are compared in fig.
\ref{fig1}. Clearly, there is poor agreement. In particular
$s/T^3$ is constant in the model as a result of the underlying AdS
geometry and $e/T^4$ is a decreasing function unlike in QCD.

The latent heat is defined as
the energy density at the phase transition. The lattice value\cite{Lucini} is 
$L_h=(0.77T_c)^4$. From (\ref{esHW}) one
finds a finite latent heat also in the HW model, $L_h  = 8 (M_p\ell)^3 \La^4 = (0.97T_c)^4$. As this is a finite quantity, {\em the transition is
of first order}. 

Presence of a first order deconfinement transition is in accord
with our expectations from large $N_c$ QCD \cite{Lucini}. However,
there are other shortcomings of the model.  First, the conformal
anomaly $T_{\m}^{\m}$ is a non-trivial function of T in QCD, see
fig. \ref{fig3} whereas this functional dependence is lost in the
hard-wall model. One can compute this as $T_{\m}^{\m}/(V_3 N_c^2)
= e_{HW} - 3 p_{HW} $ from (\ref{freeHW}) and (\ref{esHW}) and one
finds a constant $T_{\m}^{\m} = 8 N_c^2 V_3 (M_p\ell)^3
\Lambda^4$. Similarly, the speed of sound can be computed as
$c_s^2 = s/c_v$ where $c_v = d e/dT$ is the specific heat of the
system, and one finds $c_s^2 = 1/3$. This, of course reflects the
fact that the underlying geometry is AdS, and is in complete
disagreement with QCD where $c_s$ is again expected to be a
non-trivial function of T, see fig. \ref{fig3}.

Secondly, when one computes the bulk viscosity from the Kubo's formula
(see \cite{Gubser3} for a recent treatment) one finds that
$\zeta/s = 0$ which is again in disagreement with QCD. This latter
result is rather disappointing because $\zeta/s$ is considered to
be an important observable probing the quark-gluon plasma at RHIC,
and its profile as a function of T reveals important information
regarding the nature of the phase transition. In particular, both from the 
low energy theorems and lattice studies \cite{Viscos}, it is
expected to make a peak near $T_c$. 

Although there are numerous shortcomings of the the HW model\footnote{The
consistency of the model is also questionable. As
discussed in \cite{Herzog} and motivated by the critical
string-theory constructions such as \cite{KS}, the IR brane in the
AdS cavity is viewed as an ``end of space-time'' as opposed to a
boundary. However, from the point of 5D Einstein gravity, the IR
brane really acts as a boundary of the geometry in the deep
interior. Thus, in principle one should allow for a
Gibbons-Hawking term also at the location of the IR brane. The
authors of \cite{Evans} investigated this issue and found that the
deconfinement transition goes away, once a Gibbons-Hawking term is
added at the IR brane.}, it should be viewed as  a
first step in a holographic approach, instead of a
rigorous construction. Indeed, even the fact that such a simple
model captures certain basic aspects of QCD is astonishing and
should be taken as a starting point for a deeper investigation.

\begin{figure}[h]
\centerline{\psfig{file=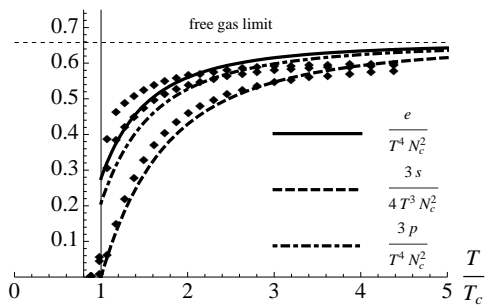,width=2.5in}} \vspace*{8pt}
\caption{Comparison of the energy, entropy and pressure densities
in the SW model with the lattice data of Boyd et
al. The diamonds refer to lattice data.\protect}\label{fig2}
\end{figure}

\subsection{Soft-wall model}\lab{SW}

Motivated by the partial success of the HW modelin the
meson sector, Karch et al. introduced an improvement in
\cite{Karch} that softens the breaking of conformality in the IR.
This is achieved by replacing the hard-wall at $r_0$ by a
non-trivial dilaton profile, \be\lab{dilSW}\phi(r) =
(\La r)^2.\ee This introduces a dimensionful parameter $\La$ that
sets the scale of the problem in the IR. The geometry is still
taken to be $AdS_5$.

One nice feature of the soft-wall model is linear confinement: The
meson spectrum is linear for large orbital excitation number and
for large spin, as opposed to the quadratic spectrum of the
hard-wall\cite{Karch}. However, some of  the unphysical features
in the HW carry over in the glue sector. In particular, there is no running gauge
coupling\footnote{One my think of $\f(r)$ as a holographic dual to
a running coupling, but this identification is problematic for
non-dynamical fields.}, and  magnetic quarks are confined. 

\begin{figure}[h]
\centerline{\psfig{file=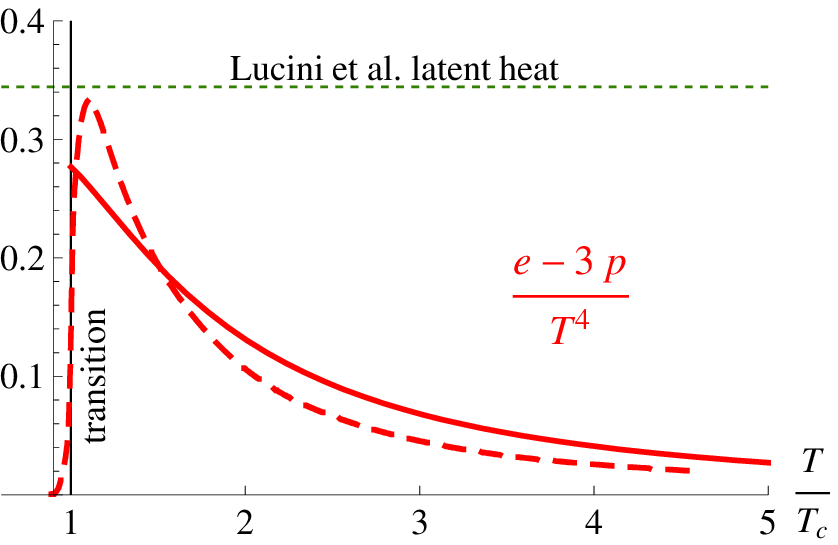,width=2.2in}\hspace{0.5in}\psfig{file=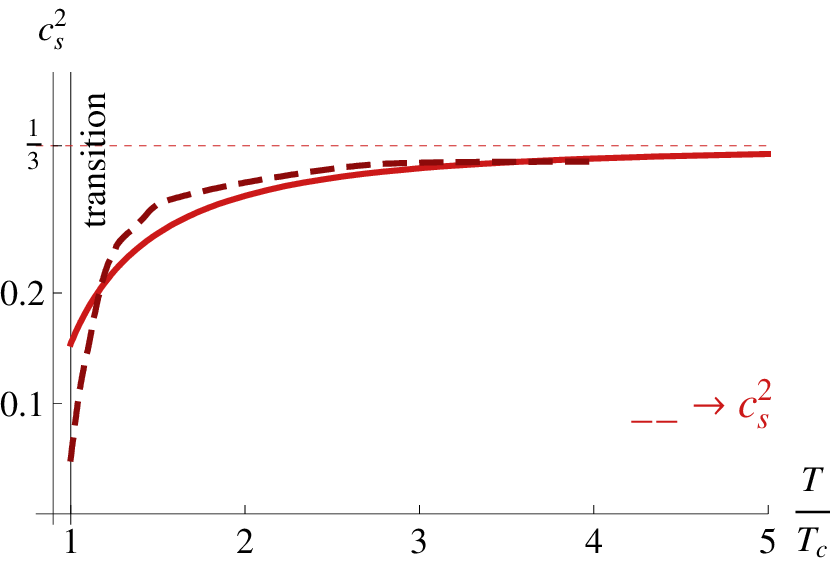,width=2.2in}}
\vspace*{8pt} \caption{Comparison of the trace anomaly and the speed of light in the SW
model with the lattice data of Boyd et al. (dashed curves).\protect}\label{fig3}
\end{figure}

Another main issue is that, the model is non-dynamical, \ie it does not follow from a 5D
gravitational action. Instead, the metric and the dilaton profile
are imposed by hand\footnote{In \cite{Gherghetta} a dynamical
Einstein-dilaton-tachyon theory is constructed that admits SW as a
solution. However it is hard to understand the presence of Tachyon
both in the gauge theory and in gravity.}. Related to this,
computation of the glueball spectra from the bulk-fluctuations is
ill defined. Below, we shall see other problematic features at
finite T\footnote{The authors of \cite{Karch} did not intend to apply the model to the 
glue sector. As a phenomenological model designed to describe the meson physics in the quenched
approximation it is indeed appropriate and the question of whether it solves the equations of motion 
is not crucial. As we discuss below, it becomes crucial when applied to thermodynamics of glue.}.

\subsection{Thermodynamics of the Soft-Wall}\lab{thermoSW}

The study of thermodynamics on this background is initiated in
\cite{Herzog}. Once again, one considers two competing solutions
at finite T: (i) the SW geometry with compact Euclidean time. (ii)
AdS black-hole, appended with the non-trivial dilaton profile
(\ref{dilSW}). As already mentioned, the construction is
non-dynamical, hence neither of these two geometries solve the
equations of motion of a 5D Einstein-dilaton system. One assumes
that they are solutions to some unspecified gravitational theory
and computes the free energy density with the prescription
described in section \ref{thermoHW} \cite{Herzog}. The result is,
\be\lab{freeSW}{\tiny\!\!\! f_{SW} = 2(M_p\ell)^3 T^4 \le[ \half+
e^{-\le(\frac{\La}{\pi T}\ri)^2}\le(\le(\frac{\La}{\pi
T}\ri)^2-1\ri)2 +\le(\frac{\La}{\pi
T}\ri)^4\!Ei[-\le(\frac{\La}{\pi T}\ri)^2]\ri].} \ee Here $\La$ is
the parameter that appears in (\ref{dilSW}) and $Ei$ is the
exponential-integral function. One obtains a phase transition at
$T_c = 0.4917\La$. As before, one can fix the value of $\La$ by
matching the lowest $\rho$ meson mass and one finds, $\La= 338$ MeV
which yields a $T_c$ better than HW, $T_c = 191$ MeV
\cite{Karsch}. The latent heat also turns out better than the HW
model. One finds $L_h = (0.725T_c)^4$ that is very close to the
Lucini et al.'s lattice result of $(0.77T_c)$.

The non-dynamical feature of the model manifests itself in the
computation of entropy. The entropy as computed from the
Bekenstein-Hawking formula and from $s=-df/dT$ above do not match.
The BH geometry does not obey the laws of thermodynamics, which
makes the findings questionable. However, let us press on, and
assume that one indeed obtains a free energy of the form
(\ref{freeSW}) from some unspecified dynamical theory and work out
other thermodynamic functions. The computation is just as in the
HW case and the results are summarized in figs. \ref{fig2} and
\ref{fig3}. These results are in very good
agreement with the lattice study of \cite{Boyd}.

It is surprising that a non-dynamical theory, constructed with
many assumptions yield such good results and it begs for a better
understanding. 
We shall, in the next section, investigate a {\em dynamical}
dilaton-Einstein system with solutions similar to the form (\ref{dilSW})
in the large $r$ region. 

An underlying dynamical theory is needed also to compute certain
important observables such as the bulk viscosity $\zeta$. In the
holographic set-up, this quantity is computed using Kubo's formula
\cite{Gubser3}. The reason this computation is ill-defined in the
SW model is that one needs to solve for the bulk fluctuations in
an holographic computation and this requires that the
background solves Einstein's eqs.

\section{Non-critical holographic QCD}

\subsection{Dynamical Models}

There is a long history of the dilaton-gravity systems in
the context of the AdS/CFT correspondence. Due to lack of space,
we are not able to provide an exhaustive list of references here.
Rather we shall mention a few articles that are closely related to
our approach. The papers \cite{Type0} 
demonstrated that type 0 string theory provides a
fruitful set-up for gravity duals of running gauge coupling. They
considered a 10D background that involves a dilaton and a bulk
tachyon field. Asymptotics of the dilaton in the deep interior
exhibits a log-running of the gauge coupling! (however, presence
of the tachyon is confusing as there are no obvious dual gauge
invariant in the gauge theory). Similarly, Gubser \cite{DF}
analyzed a dilaton flow in the context of type IIB, truncated to
5D\footnote{An early work on dilaton flow in the IIB set-up is
\cite{TypeII}}.
Other notable papers that study a dynamical dilaton flow in the 5D
set-up are \cite{NojiriOdintsov}, \cite{Gherghetta}, \cite{Evans}
and \cite{CsakiReece}. The latter uses an approach very similar to
ours\footnote{See however below eq. (\ref{IRasympii}) for various
differences.}.
Finally, Gubser and collaborators \cite{Gubser1,Gubser2} recently
analyzed the dilaton-gravity system at finite temperature,
obtaining results that are quite similar to \cite{GKMN1}.

\subsection{Improved holographic QCD}

There are many reasons supporting a non-fermionic (such as type
0 string theories) and a non-critical holographic
approach\cite{Polyakov}. From an economic point of view, five
dimensions provide all the necessary degrees of freedom to
construct a dual of QCD: four dimensions where the gauge theory
lives, plus a radial direction dual to the energy scale of the
gauge theory. Furthermore, a brief study of the low energy degrees
of freedom of 5D non-critical string theory yields a nice
correspondence between the various objects in string theory and
gauge theory \cite{GK}. Absence of extra dimensions, hence absence
of the undesired Kaluza-Klein degrees of freedom is another
attractive feature.

The only non-trivial bulk fields required to model the low energy
dynamics of large $N_c$ QCD are the metric (dual the the energy-momentum
tensor), the dilaton (dual to $\l_{YM}$ and $\tr F^2$) and
the axion\footnote{We will not be concerned with the axion in this
paper. Its action is suppressed in the large $N_c$ limit, hence can be
consistently ignored. Note however, that the axion sector has very interesting
implications for the strong CP violation problem \cite{GKN}.}
(dual to $\theta_{YM}$ and $\tr F\wedge F$). Here, we shall
present such a set-up \cite{GK,GKN} and describe its zero
temperature solutions. A simple 5D action is,
\begin{equation}
   S_5=-M^3_P N_c^2\int d^5x\sqrt{g}
\left[R-{4\over 3}(\partial\Phi)^2+V(\Phi)
\right]+2M^3_pN_c^2\int_{\partial M}d^4x \sqrt{h}~K.
 \label{act1}\end{equation}
where $V$ is a yet undetermined potential for the dilaton. The
second term above is the Gibbons-Hawking term, $K$ being the
extrinsic curvature on the boundary\footnote{As a boundary term, it has no
contribution to the equations of motion and will play no role in
this subsection. However, its contribution is crucial in comparing
on-shell actions as we discuss in the next subsection.}. 

We make the domain-wall ansatz in order to preserve the 4D Lorentz
symmetry. In the conformal coordinate system, \be\label{geo1} ds^2
= e^{2A_0(r)} \left(dr^2 + \eta_{ij} dx^idx^j\right), \qquad \Phi
= \Phi_0(r). \ee Here, $r\geq 0$ is the radial coordinate.
Boundary is located at $r=0$.

The only non-trivial input in (\ref{act1}) is the dilaton potential
$V$. In order to fix $V$ we employ requirements from the dual
gauge theory. Holographic dictionary relates the scale factor $A$
and the dilaton $\f$ to the energy scale and the 't Hooft coupling
respectively \footnote{In the latter equation, there is an
undetermined proportionality constant $\kappa$. However it can be
set to 1 by a rescaling in the potential and all physical
observables turn out independent of this rescaling. Thus, with no
loss of generality we can choose $\l_{YM} = e^{\f}$.}: \be\lab{dict} E = e^A, \qquad \l_{YM} = \l\equiv
e^{\f}. \ee Given these
identifications, one can relate the $\b$-function of the gauge
theory to $V$ in a one to one fashion\cite{GK}. Although the shape
of $V(\l)$ is not fixed without knowledge of the exact gauge
theory $\beta$-function, its UV (small $\l$) and IR (large $\l$)
asymptotics can be determined.

\vspace{0.5cm}
{\em UV asymptotics}
\vspace{0.5cm}

In the UV, the input comes from perturbative QCD. We demand
asymptotic freedom with logarithmic running. This implies in
particular that the asymptotic UV geometry is that of $AdS_5$ with
logarithmic corrections. This requires a (weak-coupling) expansion
of $V(\l)$ of the form $V(\l) = 12/\ell^2 (1 + v_1 \l + v_2 \l^2
+\cdots) $.
 Here
$\ell$ is the AdS radius and $v_i$ are dimensionless parameters of
the potential directly related to the perturbative
$\beta$-function coefficients of QCD \cite{GK}. In conformal
coordinates, close to the  $AdS_5$ boundary at $r=0$,  the metric
and dilaton behave  as \footnote{We will use a ``zero'' subscript
to indicate quantities evaluated at zero temperature.}:
\be\label{sol0UV}
    ds^2_0 = \frac{\ell^2}{r^2} \le(1+\frac89\frac{1}{\log r\La}+\cdots\ri)\le(dr^2+dx_4^2\ri), \qquad
    \l_0 = -\frac{1}{\log r\La}+ \cdots
\ee where the ellipsis represent higher order corrections that
arise from second and higher-order terms in the $\beta$-function.
The mass scale $\La$ is an initial condition for the dilaton
equation and corresponds to $\Lambda_{QCD}$ just like in the
soft-wall model above.

\vspace{0.5cm}
{\em IR asymptotics}
\vspace{0.5cm}

For any asymptotically AdS space, Einstein's equations dictate the geometry in the deep interior be,
either another AdS or a singular geometry that terminates at $r=r_0$\cite{GKN}.
For QCD,  it is the second option that is more plausible, as the gauge theory is not conformal invariant in the IR.
Details of the IR geometry (or equivalently the large $\l$ asymptotics of $V$)
 are determined by the requirement of color confinement a la \cite{Wilson}. In particular, we require that the
 quark-antiquark potential is linear.  This happens when
 \be\lab{IRasymp1}
 V(\l)\to \l^Q\log^P(\l), \qquad \l\to\infty,
 \ee
 and when the parameters $Q$ and $P$ fall into either of the two cases:
 \bea\lab{IRasympi}
 (i) && Q>4/3,\, P\,\, arbitrary\,\,\, \rightarrow\,\,\, r_0=\, finite\\
 (ii) && Q=4/3,\, P\geq 0\,\,\, \rightarrow\,\,\, r_0=\infty\lab{IRasympii}.
 \eea
 In the first case, the geometry terminates at finite $r$, hence this is somewhat similar to the hard-wall geometry.
 The latter case similar to the soft-wall geometry as it involves a singularity at $r=\infty$.
 The asymptotics above also guarantee that the classical string configurations do not reach the singularity at $r_0$.
 The same requirement for the particle-like bulk excitations yield an additional condition in case (i): $Q<4\sqrt{2}/3$.
 For $Q\geq 4\sqrt{2}/3$, the glueball spectrum becomes ill-defined\cite{GKN}.
 We note that both the geometry in \cite{CsakiReece} and in \cite{Gubser1} fall into this problematic class. These problematic
 geometries also have undesired features at finite T.

 It is shown in \cite{GKN} that
 in both cases, the magnetic quarks are screened and the glueball spectrum is gapped and discrete.
 In case (i) the glueball spectrum turns out to be quadratic whereas in case (ii) the spectrum grows as $w_n\propto n^{2P}$.
For phenomenological reasons, the preferred geometry thus corresponds to the linear spectrum with $P=1/2$. In this case the
 asymptotic geometry is,
\begin{equation}\label{sol0IR}
    ds^2_0 \to e^{- C \le(\frac{r}{\ell}\ri)^{2}}\!\!\le(dr^2+dx_4^2\ri),
    \quad\l_0 \to e^{3C/2 \le(\frac{r}{\ell}\ri)^{2}}\!\!\le(\frac{r}{\ell}\ri)^{\frac34}
\end{equation}
      where the constant $C$ is a positive constant related to $\La$ in (\ref{sol0UV}).

\vspace{0.5cm}
{\em Parameters of the model}
\vspace{0.5cm}

The dimensionless parameters of the holographic model a priori are (in AdS length units): the Planck
mass $M_p\ell$,  which governs the scale of interactions between the
glueballs in the theory, the scale $\La\ell$
that plays the role of $\Lambda_{QCD}$, the string length scale $\ell_s/\ell$ and
the parameters $v_i$ that specify the shape of the potential $V$. The Planck mass is fixed by
studying large T asymptotics, exactly as in eq. (\ref{planck}).
On the other hand, symmetries of the equations guarantee that
no physical observable depend on $\La\ell$. The number $\ell/\ell_s$ can be determined by comparison
with the string tension in lattice QCD. For the particular model that is investigated here,
this turns out to be $\ell/\ell_s\approx 8$. This is an encouraging result which shows the $\alpha'$ corrections
are suppressed by about order 10.

Finally, we fix the shape of the potential by arbitrarily picking up a
function $V$ that satisfies the UV and the IR asymptotics discussed above. A function that does the
job is,
\be\lab{pot}
V(\l) =  \frac{12}{\ell^2}\le(1+ v_1 \l + v_2 \l^{\frac43}\log^{\half}\le(1+ v_3\l^{\frac43} + v_4\l^2\ri)\ri).
\ee
We shall specify the numbers $v_i$ in the following. 

The units in the problem can be fixed by matching the lowest lying $0^{++}$ glueball in our model and
in lattice QCD\footnote{According to \cite{Meyer1} this is $m_{0++} = 1475$ MeV} \cite{GKN}.
This also fixes the the actual value of the Planck scale, $M_p N_c^{2/3}$. If one wants to compare results of the model
with a  gauge theory with finite $N_c$, this value gives a
cut-off, above which one cannot ignore string interactions. For $N_c=3$ one has $M_p N_c^{2/3}\approx 2.5$ GeV.
Of course there is no such a cut-off in large $N_c$ QCD.

\section{Thermodynamics of Improved Holographic QCD}

Having defined the theory at zero T, now we look for finite T
solutions. At finite temperature
there exist two distinct types of solutions to the action
(\ref{act1}) with AdS asymptotics, (\ref{sol0UV}):
\begin{enumerate}
  \item[i.] The thermal graviton gas, obtained by compactifying
  the Euclidean time in the zero temperature solution with  $\tau\sim \tau+1/T$ :
\begin{equation}\label{TG}
    ds^2 = e^{2A_0(r)}\le(dr^2 + d\tau^2+ dx^2_3\ri), \,\, \l=\l_0(r).
\end{equation}
This solution exists for all $T\geq 0$ and corresponds to a confined phase,
if the gauge theory at zero T confines.

\item[ii.] The black hole (BH) solutions (in Euclidean time) of the form:
\begin{equation}\label{BH}
    ds^2 = e^{2A(r)}\le(\frac{dr^2}{f(r)} + f(r) d\tau^2+ dx^2_3\ri), \,\, \l=\l(r).
\end{equation}
The function $f(r)$ approaches unity close to the boundary at
$r=0$. There exists a singularity in the interior at $r=\infty$
that is now  hidden by a regular horizon at $r=r_h$ where $f$
vanishes. Such  solutions correspond to a  deconfined phase.
\end{enumerate}

As we discuss below, in confining theories the BH solutions exist
only above a certain minimum temperature, $T>T_{min}$.

The thermal gas solution has only two parameters: T and $\La$. The
black hole solution should also have a similar set of parameters:
 the equations of motion are
second order for $\l$ and $f$, and first order for $A$
\cite{GKMN2}. Thus, {\em a priori} there are 5 integration
constants to be specified. A combination of two integration
constants of $A$ and $\l$ determines $\La$. (The other combination
can be removed by reparametrization invariance in $r$). The
condition $f\to 1$ on the boundary removes one integration
constant and demanding regularity at the horizon, $r=r_h$, in the
form $f\to f_h(r_h-r)$, removes another. The remaining integration
constant can be taken as $f_h$ (or $r_h$, they are not
independent), related to the temperature by \be\lab{T}4\pi T=
f_h.\ee

In the large $N_c$ limit, the saddle point of the action is dominated by one of the two types of solutions.
In order to determine the one with minimum free energy, we
need to compare the actions evaluated on solutions i. and ii. with equal
temperature.

We introduce a cutoff boundary at $r=\eps$ in order to regulate
the infinite volume. The difference of the two scale factors is
given near the boundary as \cite{GKMN2}:
\begin{equation}\label{bb0}
    A(\eps) - A_0(\eps) = \mathcal{G}(T) (\eps\La)^4+\cdots
\end{equation}
Then the free energy density is given by \cite{GKMN1}:
\be\lab{freeNC} f_{NC} = -p_{NC} = 15 (M_p\ell)^3 \La^4
\mathcal{G}(T)-{T s_{NC} \over 4}. \ee Here, the entropy density
$s_{NC}$ is given by the area of the horizon: \be\lab{sNC} s_{NC}
= 4\pi^2 M_p^3 e^{3A(r_h)}. \ee {\em One can check (by numerics)
that this entropy is precisely the same as follows from the 1st
law $s= -df/dT$.} This is what one expects as the theory defined
by (\ref{act1}) satisfies the gravitational energy theorems and
$T$ is defined in (\ref{T}) by requiring absence of conical
singularity at the horizon.

It is clear from (\ref{freeNC}) that presence of the first term is
crucial for existence of a phase transition, as the second term by
itself is negative definite. Below, we explain the physical
meaning of the quantity $\cG$.

\vspace{0.5cm} {\em Role of the gluon condensate}
\vspace{0.5cm}

The quantity $\cG$ can also be defined from the difference of the
dilatons ($\l=\exp(\f)$), \be\lab{difdil} \F(\eps)-\F_0(\eps) = \frac{45}{8}
\cG(T)(\eps\La)^4\log(\eps\La)+\cdots \ee Now, the meaning of
$\cG$ becomes clear. The AdS/CFT prescription relates bulk
fluctuations with VeVs of dual operators in the gauge theory. As
the dilaton couples to the operator $\tr F^2$, we learn that $\cG$
is the difference of VEVs in the gluon condensate at finite and
zero temperatures: $\mathcal{C}(T) \propto\la\tr F^2\ra_T - \la\tr
F^2\ra_0 $.

Let us perform a consistency check. The dilatation Ward identity
in gauge theory relates the condensate to the energy-momentum
tensor: $ \la T_{\m}^{\m}\ra_{T0} =
-\frac{\b}{4\l^2}~\la\tr F^2\ra_{T0} $. The subscript refers to
the difference finite and zero T. We shall check this identity in
the holographic set-up (at leading order in $\l$). The LHS follows
from $T_{\m}^{\m} = \e_{NC} - 3p_{NC}$. The energy $e_{NC}$ is
derived from (\ref{act1}), one finds \be\lab{conds} T_{\m}^{\m}
=60 (M_p\ell)^3\La^4 \cG. \ee

The RHS of the Ward identity is computed by the AdS/CFT
prescription: For any canonically normalized bulk fluctuation for
$\chi(x) = r^{\Delta_-}\chi_0(x)+r^{\Delta_+}\chi_1(x)$ near the
boundary, the VeV of the dual operator is $\la\cO(x)\ra =
(2\Delta_+ - d)\chi_1(x)$. Taking $\chi$ as $\delta\F$ in
(\ref{difdil})\footnote{One should be careful about the
multiplicative factors arising from normalization of $\F$ in
(\ref{act1}), see \cite{GKMN2}.}, we find $\la \tr F^2\ra_{T0} =
\frac{240(M_p\ell)^3N_c^2\La^4}{b_0} \cG$. Using the $\b$-function
$\b(\l)= - b_0\l^2-\cdots$ we see that this precisely matches the
RHS of the Ward identity given by (\ref{conds}).

One may wonder why it works. After all, the prescription is
conjectured for the pure AdS space and we have a log-corrected AdS
here. The reason is that, one can generalize the holographic
renormalization program of AdS to this geometry by explicitly
computing the counter-terms \cite{GKP} and show that the
contribution from the counter-terms cancel out precisely between
the finite and zero T components.

\vspace{0.5cm} {\em Existence and order of the deconfinement
transition}
\vspace{0.5cm}

For a general potential $V$ that obeys the UV and IR asymptotics
described in the previous subsection, we can prove the following
statements:

\begin{itemize} \item[i.] {\em There exists a
phase transition at finite T, if and only if the zero-T theory
confines as in (\ref{IRasympi}) or (\ref{IRasympii})} 
\item[ii.]
{\em This transition is of the {\bf first order} for {\bf all} of
the confining geometries, with a single exception described in
iii:}
\item[iii.] {\em In the limit
confining geometry $P=0$ of (\ref{IRasympii}), $A_0(r)\to -C r$
(as $r\to \infty$), the phase transition is of the {\bf second
order} and happens at $T = 3C/4\pi$.}
\item[iv.] {\em All of the non-confining geometries at zero T are always in the
black hole phase at finite T. They exhibit a second order phase
transition at $T=0^+$.}
\end{itemize}
An heuristic demonstration is given in \cite{GKMN1} and a general,
coordinate independent proof will appear in \cite{GKMN2}. Here,
let us only mention that the crucial element for the phase
transition in confining geometries is the existence of (i) a
``big" black-hole with positive specific heat for small $r_h$ and
(ii) a ``small" black-hole with negative specific heat for large
$r_h$. Co-existence of big and small black-holes is just as in AdS
BHs {\em with spherical horizon}. See fig. \ref{illus} for an
illustration. It is clear from this figure that there exists a
$T_{min}$ for the confining geometries as in eq. (\ref{IRasympii}, 
below which both BHs disappear.

\begin{figure}[h]\lab{illus}
\centerline{\psfig{file=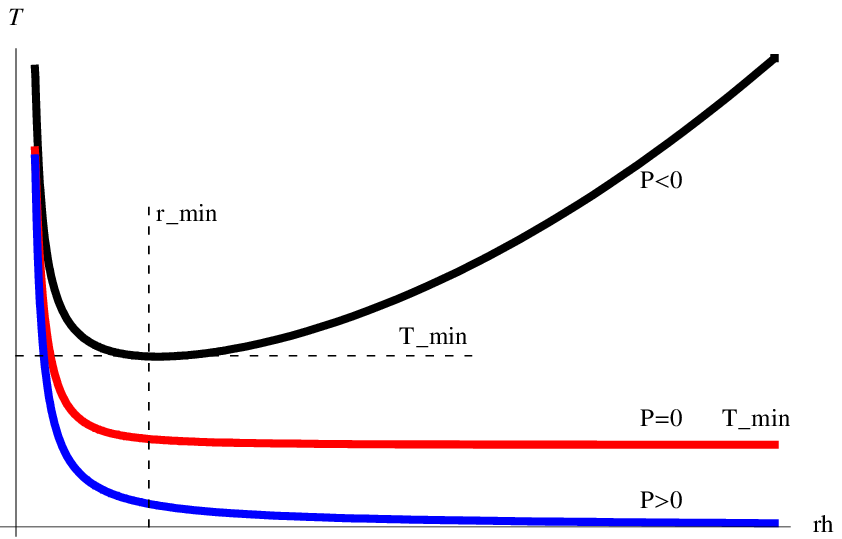,width=2.2in}\hspace{0.5in}\psfig{file=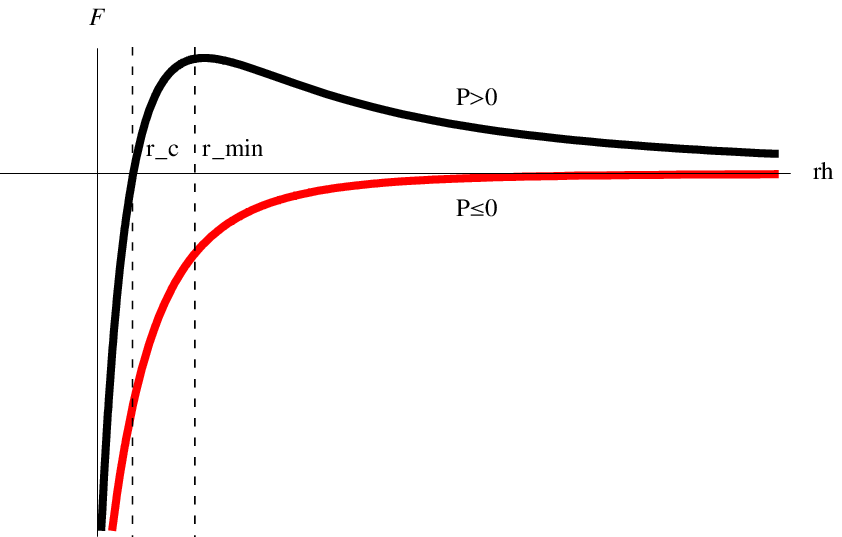,width=2.2in}}
\vspace*{8pt} \caption{Schematic behavior of temperature and
free energy as functions of $r_h$, for the
infinite-$r$ geometries of the
 type (\ref{IRasympii}), for different values of $P$.}
\end{figure}

\vspace{0.5cm} {\em Numerical Results}
\vspace{0.5cm}

The numerical results that we review in this section are based on 
\cite{GKMN3}.  
All the thermodynamic properties of the system follow from
(\ref{freeNC}). One numerically solves the Einstein-dilaton system
for a fixed $\La$\footnote{This is the same in both geometries and
fixed by the lowest $0^{++}$ mass as $\La\approx 290$ MeV.} and
for different $r_h$, corresponding to different T (see fig.
\ref{illus}) to obtain $s_{NC}(T)$ and $\cG(T)$. The rest
follows from the laws of thermodynamics. The potential is chosen
in (\ref{pot}). Only three of the $v_i$ are independent because,
as mentioned earlier, the physics is left invariant under the
rescaling $\l\to \kappa\l$. We fix one combination of $v_i$ to
match the lattice result for the latent heat $L_h = (0.75 T_c)^4$.
The two other parameters are chosen in order to obtain good
glueball mass ratios\footnote{We shall not discuss the glueball
spectrum here, see \cite{GKMN3}. With the potential above, one
obtains e.g.  $m_{0++*}/m_{0++} = 1.6$ which is in well agreement
with lattice \cite{Meyer1}.}. A good set of parameters is
$\{v_1,v_2,v_3,v_4\} = \{0.1,46,0.05,1000\}$.\footnote{The difference in these coefficients and the 
ones in \cite{GKMN3} are due to a different choice of $\kappa$ here.}
The rest of the results in this section are predictions.

\begin{figure}[h]
\centerline{\psfig{file=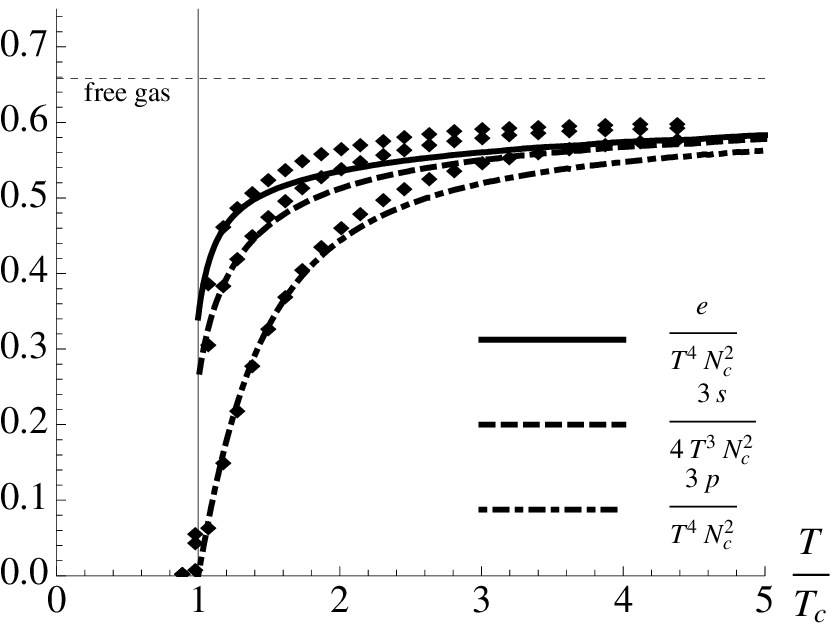,width=2.2in}\hspace{0.5in}\psfig{file=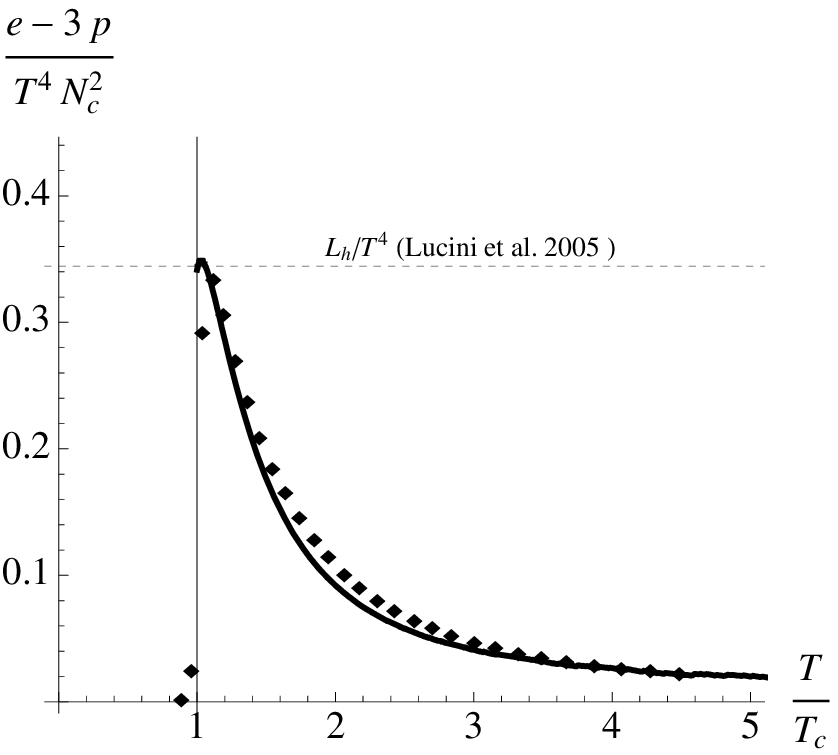,width=2.2in}}
\vspace*{8pt} \caption{Dimensionless thermodynamic functions
and the gluon condensate. The diamonds correspond to the lattice
 data of Boyd et al.\protect}\label{fig5}
\end{figure}

\begin{figure}[h]
\centerline{\psfig{file=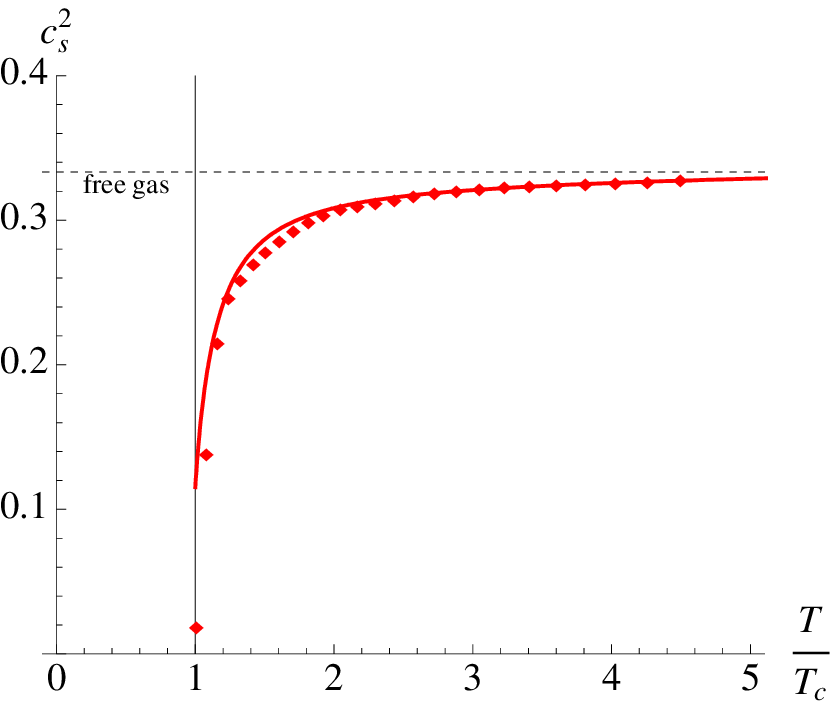,width=2.2in}
\hspace{0.5in}\psfig{file=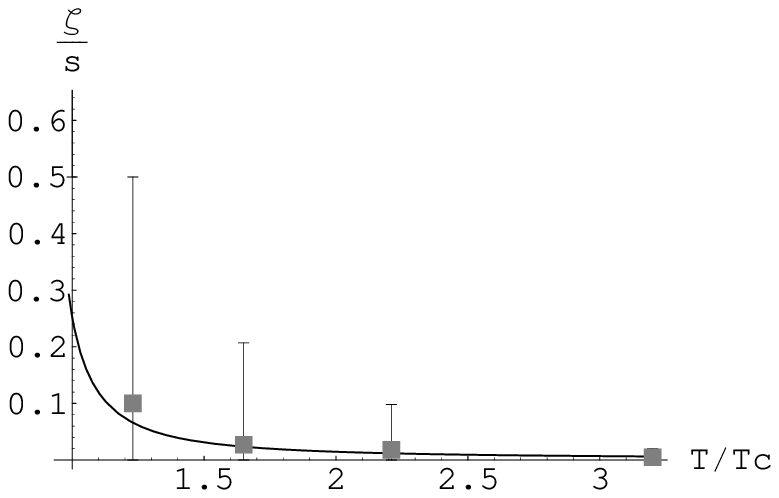,width=2.2in}}
\vspace*{8pt} \caption{Left: Comparison of speed of sound in our model
and the lattice result of Boyd et al. (diamonds). Right: Comparison of the bulk viscosity wit the lattice data of Meyer. \protect}\label{fig6}
\end{figure}

We find a transition temperature at $T_c\approx 247$ MeV which is very close to lattice
\cite{Lucini}.  
\footnote{This value is for $SU(N_c)$ YM in the large $N_c$ limit 
which is significantly different from both the QCD value and the SW model 
cited in sec. \ref{thermoSW}.} The
thermodynamic functions $\e_{nc}$, $s_{nc}$ and $p_{nc}$ are
compared with the lattice data in fig. \ref{fig5}, left. The
temperature dependence of the gluon condensate is shown and
compared to lattice in fig. \ref{fig5}, right. The speed of sound
and the bulk viscosity are presented in fig. \ref{fig6}.\footnote{The derivation of numerical 
results on the bulk viscosity, along with other dynamical observables  will appear in \cite{GKMiN}.}
We conclude that the model presented here is in very good agreement with the
available lattice data.

A last word on the bulk viscosity. Both the low-energy theorems
and the lattice arguments \cite{Viscos} indicate that the bulk
viscosity has a peak near $T_c$. This is what we also observe in
fig. \ref{fig6}, however the height of the peak is less than the
lattice evaluation \cite{Meyer2}\footnote{Note however that
lcomputation of this quantity on the lattice is notoriously
difficult and afflicted with numerical errors that arise from
analytic continuation.}, see also \cite{Gubser3} for the same
conclusion.

\section{Discussion and Outlook}

We presented a holographic model for large $N_c$ QCD at finite T, 
that resolves most of the problematic issues of the AdS/QCD models
and yields very good agreement with the available lattice data. The deconfinement 
transition results from presence of a non-trivial gluon condensate. We also demonstrated
that the AdS/CFT prescription for computing n-point functions carry over if computed 
as differences at finite and zero T. Strictly speaking, the model is valid at large $N_c$. 
For finite $N_c$, there exists a UV cut-off, which is about $2.5\, GeV$ for $N_c=3$. 
The $\alpha'$ corrections are somewhat under control as the AdS radius is order 10 in string units. However, generally 
one expects corrections from the higher string modes. 

One related problem of all two-derivative effective actions is that the shear viscosity - entropy ratio 
is universally fixed as $\eta/s = 1/4\pi$ \cite{Buchel}, rather than a function of T as expected in QCD. In order to 
cure this problem, one should consider higher derivative corrections in the action. 
Other possible future directions include study of the meson sector via probe D4 branes, turning on a baryon 
chemical potential by charged BHs and eventually searching for explicit non-critical or critical string theory 
backgrounds where the solutions can be embedded. 

\section*{Acknowledgments}
Some of the results on the thermodynamics of the HW and SW models,
and all of the results in sec. 3 were derived in collaboration
with E. Kiritsis, L. Mazzanti and F. Nitti. We thank E. Kiritsis and K. Peeters for a careful reading and comments. 
This work was supported by the VIDI grant 016.069.313 from the Dutch organization for Scientific Research (NWO).

  \end{document}